# Dual-frequency injection-locked continuous-wave near-infrared laser

Trivikramarao Gavara,[1] Takeru Ohashi,[1] Yusuke Sasaki,[1] Takuya Kawashima,[1] Hiroaki Hamano,[1] Ryo Yoshizaki,[1] Yuki Fujimura,[1] Kazumichi Yoshii,[1,3] Chiaki Ohae,[2,3] and Masayuki Katsuragawa[1,2,3,*]

[1] *Graduate School of Informatics and Engineering, University of Electro-Communications, 1-5-1 Chofugaoka, Chofu, Tokyo 182-8585, Japan*
[2] *Institute for Advanced Science, University of Electro-Communications, 1-5-1 Chofugaoka, Chofu, Tokyo 182-8585, Japan*
[3] *ERATO, JST, MINOSHIMA Intelligent Optical Synthesizer Project, Honmachi 4-1-8, Kawaguchi, Saitama 332-0012, Japan*
*Corresponding author: katsuragawa@uec.ac.jp

**Abstract:** We report a dual-frequency injection-locked continuous-wave near-infrared laser. The entire system consists of a Ti:sapphire ring laser as a power oscillator, two independent diode-lasers employed as seed lasers, and a master cavity providing a frequency reference. Stable dual-frequency injection-locked oscillation is achieved with a maximum output power of 2.8 W. As fundamental performance features of this laser system, we show its single longitudinal/transverse mode characteristics and practical power stability. Furthermore, as advanced features, we demonstrate arbitrary selectivity of the two frequencies and flexible control of their relative powers by simply manipulating the seed lasers.

The advantage of dual-frequency lasers is that, in addition to their system compactness, mutual fine overlaps in time and space between radiations at two frequencies can be automatically realized, because the two-frequency radiation is produced by a single laser-resonator. Many types of dual-frequency lasers have been reported, including diode lasers with a periodic phase-change grating [1]; Ti:sapphire lasers implementing self-seeding technology or intracavity frequency-selectors, or both [2,3]; and fiber lasers with Bragg gratings and ultranarrow bandpass filters [4]. Above all, dual-frequency injection-locked (DFIL) lasers [5-8] are powerful tools, especially in the study of nonlinear optical processes in isolated atoms or molecules [9, 10]. This is because, in addition to having general advantages as dual-frequency lasers, they can widely generate a variety of two-frequency combinations that simultaneously have high spectral purity and high power. So far, these DFIL lasers have been developed on the basis of the nanosecond pulsed laser [5-8], and not yet under a continuous-wave (cw) regime. If we can extend DFIL nanosecond pulsed-laser technology to the cw regime, it will be a very attractive tool for applications such as high-precision nonlinear spectroscopy, atmospheric science, and THz wave generation. In this *Letter*, we report the development of a DFIL cw laser. We show the key performance features of this laser, which is controlled precisely at a variety of two-frequency combinations. (In regard to single-frequency injection-locked laser that is closely related to this study, see Refs. 11-14.)

A schematic of the DFIL cw laser system is illustrated in Fig. 1. The system consists of a master cavity, two seed lasers, and a power oscillator. The master cavity is a Fabry-Perot cold cavity with a finesse of 6,500 (free spectral range (FSR):

2.5 GHz) over a wavelength range of 700 to 900 nm. The incident mirror is curved (radius curvature: 200 mm) and the end-mirror mounted on a piezoelectric transducer (PZT) is flat, where the beam waist at the end mirror is 300 $\mu m$ at diameter. This master cavity provides a frequency reference for the entire laser system.

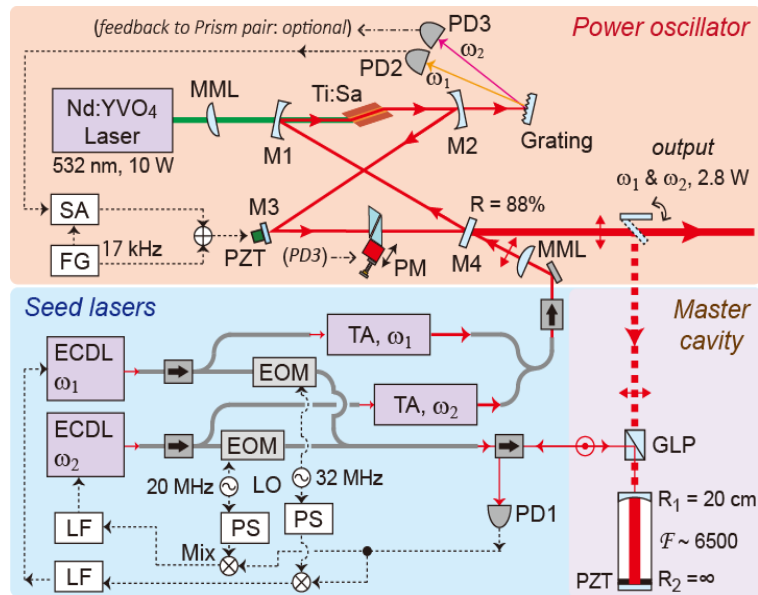

**Fig. 1. Schematic of the dual-frequency injection-locked continuous wave laser.**
ECDL, external-cavity controlled diode laser; EOM, electrooptic modulator; TA, tapered amplifier; GLP, glan laser prism; PD, photo detector; PS, phase shifter; LO, local oscillator; LF, loop filter; SA, servo amplifier; MML, mode matching lens; M, mirror; PZT, piezoelectric transducer; PM, picomotor; FG, function generator.

The two seed lasers are custom made external-cavity diode-lasers (ECDLs), the oscillation frequencies of which, $\omega_1$, $\omega_2$, are locked to the single master cavity by employing the Pound-Drever-Hall (PDH) method [15]. In order to lock the two frequencies simultaneously, the two PDH feedback loops are composed independently for each of the ECDLs $\omega_1$ and $\omega_2$, although their loops are partly in common, as illustrated. The detailed architecture of the PDH feedback-loops is as follows. Fiber dividers separate small fractions (20%) of each of the two ECDL outputs, and electrooptic modulators give phase modulations to each of them at $\omega_1$: 20 and $\omega_2$: 32 MHz. Then a fiber combiner combines the modulated radiations into a single mode fiber and introduces them into the master cavity through an isolator. The radiation reflected from the cavity is detected by a photo detector (PD1) and split into two feedback loops by using a power divider. Each of the electrical signals is then mixed with the reference RF signals at 20 and 32 MHz from function generators, producing two independent error signals for the $\omega_1$ and $\omega_2$ oscillations. Finally, the error signals are fed back to the ECDL current controllers through loop filters and lock the oscillation frequencies, $\omega_1$ and $\omega_2$, to the master cavity. After the seed radiations at $\omega_1$ and $\omega_2$ are stabilized in this way, the main part of the ECDL output (80%) are further amplified with tapered amplifiers, each up to ~100 mW, then again combined into a single mode fiber, and finally coupled to the power oscillator with a mode matching lens.

The power oscillator is constructed with a bow-tie ring cavity configuration (round-trip length: 480 mm; FSR: 625 MHz), which includes four mirrors, a gain medium, and a pair of glass wedges. Two of the four mirrors, M1 and M2, are curved, with a radius of curvature of 100 mm, creating a small spatial-mode waist (diameter at $1/e^2$: ~80 $\mu m$) between M1 and M2. The other two, M3 and M4, are flat mirrors so that a nearly flat mode profile (diameter at $1/e^2$: ~0.6 mm) is created in the other region. Mirror M2 is mounted on a translation stage to adjust the distance between the two curved mirrors, which is critical to the cavity-spatial-mode profile. Also, one of the flat mirrors, M3, is mounted on a PZT to precisely adjust and modulate the cavity length (17 kHz). The remaining flat mirror, M4, serves as an output coupler with a reflectivity of 88% (all mirrors other than M4 have reflectivities of >99%). The gain medium is a Brewster-angle-cut Ti:sapphire crystal with a length of 20 mm ($Ti^{3+}$ dopant: 0.25%), mounted on a water-cooled copper heat sink, which is placed at the spatial-mode waist between the two curved mirrors, M1 and M2. We also place one optical element other than the gain medium, namely a pair of glass wedges (borosilicate glass, vertex angle of the wedges = 30 degrees), between the two flat mirrors, M3 and M4. One of the wedges is mounted on a translation stage (New Focus, picomotor 9061-X-P-M) as illustrated [16]. This wedge pair adjusts the dispersion inside the cavity to achieve simultaneous resonance of the two seed frequencies at the power oscillator [8]. The wedge is precisely moved along its adjacent (from ~10 nm to a few mm), whereas the cavity confinement is not disturbed.

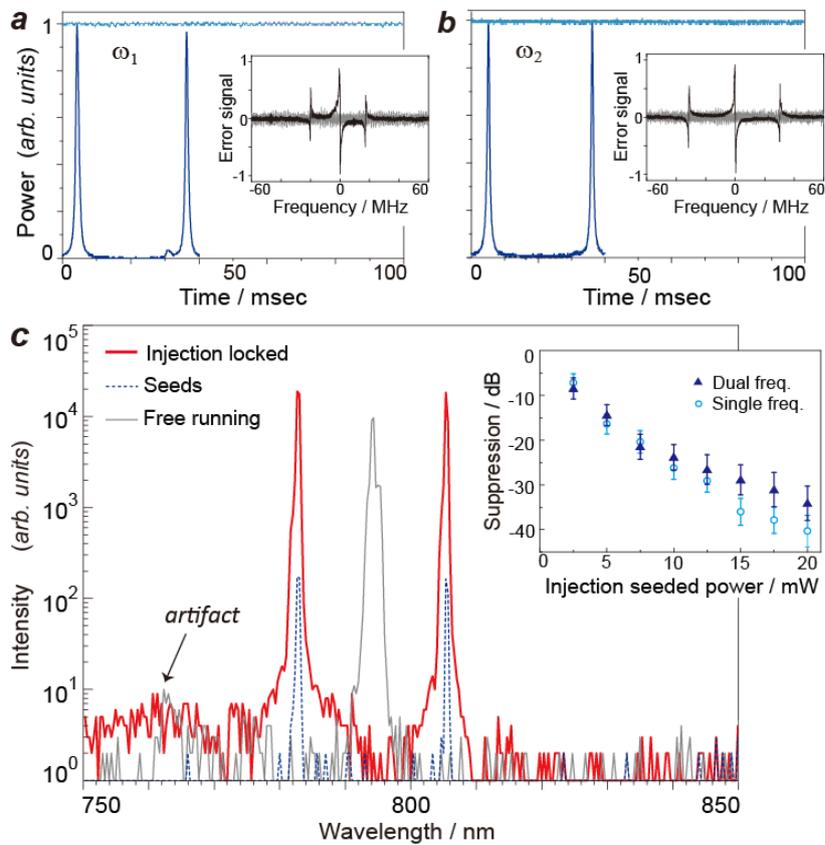

**Fig. 2. Dual-frequency injection-locked oscillation.** Seed radiation in the power oscillator at *a*, $\omega_1$ and *b*, $\omega_2$, injected under dual-resonance conditions. *c*, Spectral output from the power oscillator: injection-locked with pump (red), seed (blue), and free-running (gray).

The dual-frequency injection-locking is processed as follows. First, we check the resonance conditions of the power oscillator for each of the two seed frequencies, $\omega_1$ and $\omega_2$. For this purpose, we pick up a small leakage power from mirror M2, separate it into the individual frequencies, $\omega_1$ and $\omega_2$, by using a grating element, and detect each by using two independent photo detectors (PD2, PD3). Then, we sweep the ring-cavity length widely and monitor the longitudinal modes of the two frequencies. Next, we adjust the dispersion inside the power oscillator by referencing these two frequency modes. In other words, we adjust the wedge insertion-length precisely with a precision of ~10 nm so that the power oscillator has precisely simultaneous resonance at the two seed frequencies. Finally, we lock the power oscillator tightly into the dual-resonance condition by employing a general cavity-length modulation method, whereby mirror M3 mounted on the PZT is slightly modulated at 17 kHz. Note that, here, the power oscillator is also stabilized to the master reference cavity through one of the seed radiations, $\omega_1$.

As a pump laser, we employ a frequency-doubled Nd:YVO$_4$ laser (Coherent, Verdi-10). The 532-nm pump radiation is introduced into the power oscillator through curved mirror M1, where a mode-matching lens, MML, is tilted slightly to create an optimal spatial overlap between the pump radiation and the cavity spatial mode, including their astigmatisms.

Now we move onto the results obtained in DFIL operation. The oscillation frequencies of the ECDLs, $\omega_1$ ($\lambda_1$ = 783.8849 nm) and $\omega_2$ ($\lambda_2$ = 806.2823 nm), were locked to the master cavity by using the PDH method. These wavelengths were chosen as typical examples by referencing a wavemeter (Anritsu, optical wavelength/ frequency counter MF9630A). The respective error signals (gray dots) in the PDH locking are shown in the insets in Figs. 2a, b, together with the entire error-signal profiles (black curves). The spectral purities were estimated to be ∼40 kHz (at rms) under this PDH-locked condition. We introduced these seed radiations stabilized at $\omega_1$ and $\omega_2$, into the power oscillator. The seed powers at $\omega_1$ and $\omega_2$, sent to the power oscillator, were set at 15 mW each.

The two-frequency longitudinal modes observed with the power oscillator are shown by the blue curves in Figs. 2a, b. The seed radiations were coupled well to the power oscillator, with an efficiency of 90%. From these mode profiles, the finesse of the power oscillator was estimated to be 35, which was close to the finesse of 37 derived from the output-coupler reflectivity of 88% and one-round-trip loss in the oscillator, 4%. After we had adjusted the dispersion precisely inside the power oscillator so that we had obtained the dual-resonance condition described above, we locked the power oscillator to the seed at $\omega_1$. The blue dots in Figs. 2a, b are stabilized seed powers at $\omega_1$ and $\omega_2$ in the power oscillator, which were monitored by photo detectors, PD2 and PD3. Fluctuation was suppressed sufficiently to less than 2%.

Under this dual-resonance condition, we introduced pump laser radiation at a maximum of 10 W. We readily obtained a DFIL oscillation. Fig. 2c shows a typical output spectrum from the power oscillator, which were observed with an optical multichannel analyzer (OMA; Andor). A two-frequency spectrum (red line) appeared sharply at 784 nm and 806 nm, coinciding with the frequencies of the seed lasers (blue dots). A spectrum in the free-running oscillation (gray) is also shown here; it had a central peak at around 795 nm. This free-running radiation was output in both forward and backward directions with equal power, while they were greatly suppressed (by >30 dB) under injection-locked operation, as seen here.

To know the minimum seed-power required for DFIL operation, we studied the suppression of free-running oscillation as a function of injected seed power. The blue triangles in the inset plot in Fig. 2c indicate the result. Suppression exceeded 30 dB for a 20-mW total injected seed power. This suppression behavior for the total seed powers at the two frequencies was similar to that with a single-frequency injection-locked oscillation (open blue circles). The seed power required for stable DFIL oscillation can also be derived by estimating the locking range [17], yielding a result of ∼10 $\mu W$ for a seed linewidth of 40 kHz. As observed here, in reality 20 mW was required to have sufficient suppression of 30 dB, much larger than the above estimation [18].

In regard to the DFIL operation obtained, we next studied the fundamental performance of the entire laser system in detail. The lasing threshold was 2.0 W with regard to pump power, and the maximum output obtained was 2.8 W with a 10-W pump. The slope- and energy-conversion efficiencies were 35% and 28%, respectively. These specifications were nearly the same as those in single-frequency injection-locked operation and also in free-running oscillation (for total power in the forward and backward directions).

Figures 3a, b, and c, represent other fundamental performance characteristics of DFIL oscillation, namely *a*, spectral purity; *b*, spatial-mode purity, and *c*, power stability. Spectral purity was estimated by using optical spectrum analyzer (FSR: 2 GHz, Finesse: 100). The two longitudinal modes (red), corresponding to the two seed frequencies, $\omega_1$ and $\omega_2$, were clearly confirmed with an instrumental resolution of 20 MHz; no other modes were seen, unlike the case with the free-running oscillation (gray). We further evaluated spectral purity by creating a beat with the seed whereby the seed frequency was shifted by 110 MHz with an amplitude optical modulator. No apparent spectral broadening from the seed

linewidth was observed at a 1-kHz frequency resolution using an RF spectrum analyzer (see inset in Fig. 3a). Nearly same linewidth as those of the seeds was preserved.

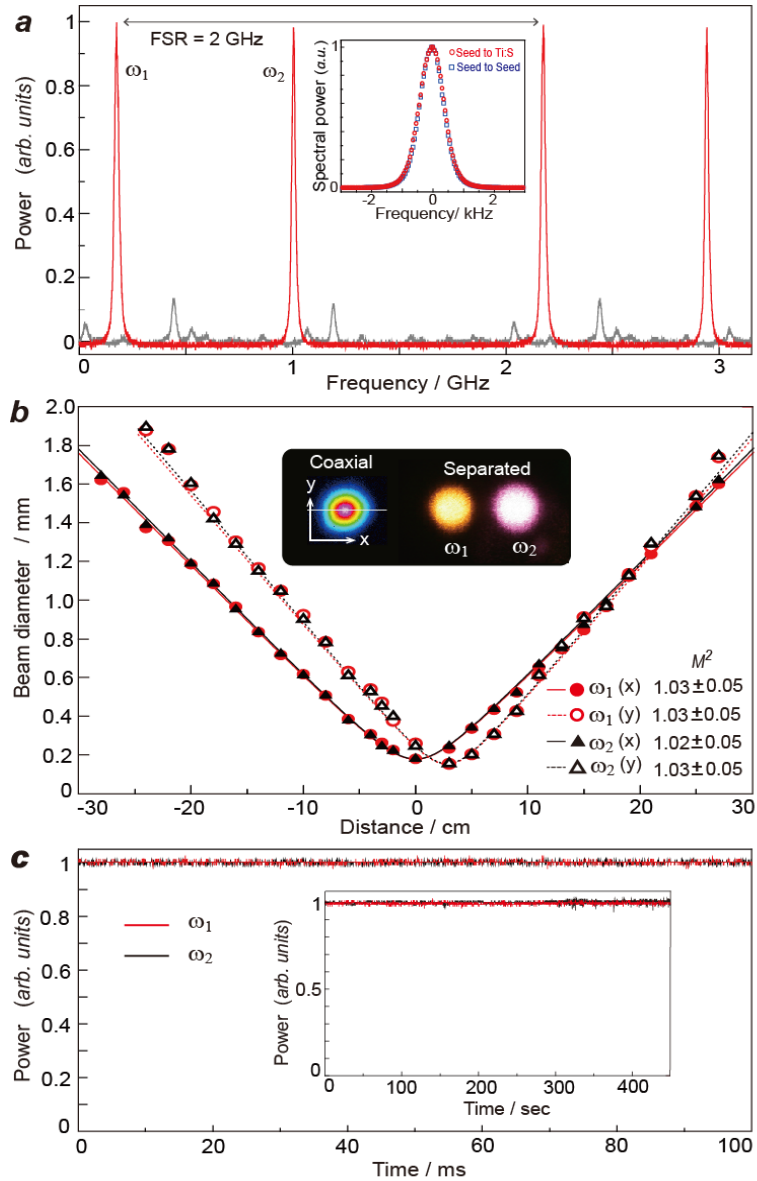

**Fig. 3. Fundamental performance characteristics of the dual-frequency injection-locked laser**: a, longitudinal mode; b, transverse mode; c, output power stability.

Spatial mode purity was measured with the $M^2$ method (Fig. 3b). We focused the DFIL output beam onto a CCD-based beam profiler (Gentec-WincamD) by using a convex lens (f: 300 mm) and measured the beam diameters at $1/e^2$ as a function of distance, where we selected either of the two frequency beams by using a bandpass filter. The beam-quality factors, $M^2$, were close to unity in both the X and Y directions for either of the two-frequency beams, as indicated, although a small astigmatism was included mainly because of the bow-tie ring cavity configuration. We note that this astigmatism can be compensated simply by using an appropriate cylindrical-lens pair, if required in an application. The inset in Fig. 3b shows the beam profile of the DFIL output and photos of the individual outputs at $\omega_1$ and $\omega_2$, taken after passage through a dispersive prism. The output beams had a perfect spatial overlap between $\omega_1$ and $\omega_2$.

Output power stability was measured simply by using photo detectors (Fig. 3c). The fluctuations were 1% at $\omega_1$ (red) and 2% at $\omega_2$ (black) on a short time scale of 100 ms; they were similar even over a long time scale of 450 s, i.e., 1% at $\omega_1$ and 2% at $\omega_2$ (see inset). Here, we did not see any instability caused by nonlinear processes competing between the two frequencies. The slightly larger fluctuation at $\omega_2$ was due to the mechanism whereby the power oscillator was locked to the master reference cavity by the $\omega_1$ seed.

Up to this point, we have described the fundamental performance characteristics of DFIL operation, namely single longitudinal/transverse mode nature and practical output-power stability. These characteristics imply that the DFIL laser is applicable to a variety of purposes.

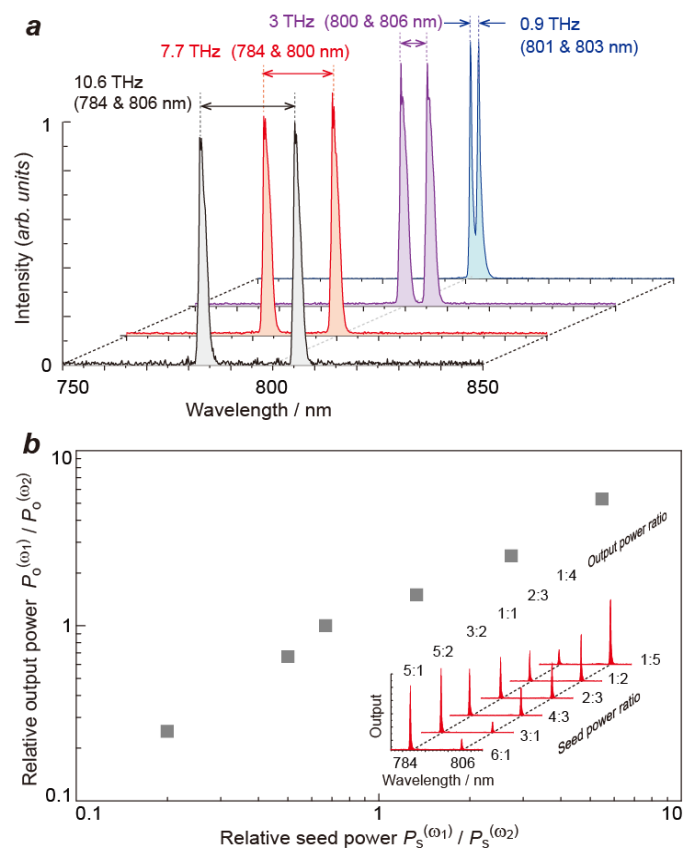

**Fig. 4. Advanced performance characteristics of the dual-frequency injection-locked laser:** *a*, selectivity of two frequencies; *b*, controllability of output-power ratios at the two frequencies. The inset DFIL output spectrum was observed with a multichannel spectrometer.

From the perspective of applications, especially when they include nonlinear optical processes, selectivity of the wavelength combinations and precise controllability of the output-power ratios at the two selected wavelengths will be key issues. Finally, we demonstrate these advanced abilities in this DFIL laser system.

We tested various frequency combinations with frequency spacings from 500 GHz to 11 THz; several of them are shown in Fig. 4a. The frequency spacings and the selected wavelengths were as follows: 10.623 THz (783.8849 nm,

806.2823 nm), 7.731 THz (783.8853 nm, 800.0595 nm), 2.895 THz (800.0590 nm, 806.2885 nm), and 0.867 THz (801.1395 nm, 803.0041 nm). For all of these wavelength-pairs, we obtained stable DFIL oscillations with performances equivalent to those shown in Figs. 2 and 3. Even larger frequency spacing is possible. If we exchange the output coupler and seed lasers appropriately, depending on the wavelength region, our DFIL laser can cover the entire gain region of a Ti:sapphire laser, namely 670 to 1050 nm [19]. Note that the two frequencies can be selected continuously, although the power oscillator has a discrete mode. For any frequency combinations, whereas their frequency spacing must be greater than 100 GHz [8], we can achieve dual resonance condition by slightly adjusting the insertion thickness of the wedge pair into the power oscillator.

Another advanced ability: control of the output-power ratio is shown in Fig. 4b. The output-power ratios at the two frequencies, $\omega_1$ and $\omega_2$, can be flexibly controlled over a wide dynamic range of greater than one order by simply manipulating the seed-power ratios at the two frequencies. At any power ratio, stable DFIL oscillations were observed with the same total output power, namely 2.8 W; ($\omega_1$ / W, $\omega_2$ / W) = (2.30, 0.47), (1.95, 0.84), (1.66, 1.12), (1.38, 1.40), (1.14, 1.62), (0.55, 2.24). Here, we preserved total seed power at a constant to maintain good free-running-suppression of <$10^{-3}$. Note that the frequencies $\omega_1$ ($\lambda_1$= 783.8849 nm) and $\omega_2$ ($\lambda_2$= 806.2823 nm) employed here, provided a calculated gain-to-loss product ratio close to unity; this implies that the output power ratio varies almost linearly with the seed power ratio [6]. This expectation was also confirmed, as seen here. When we employed other frequency combinations, the output-power ratios followed different curves, whereas the controllabilities were the same.

Before we conclude, we briefly describe an advanced application of this DFIL laser. The master cavity can also simultaneously act as an enhancement cavity for performing intracavity linear/nonlinear optics experiments (see Fig. 1) [20, 21]. Because the two seeds are locked to the master cavity in advance, the DFIL-laser output is automatically coupled to the high-finesse master cavity. A coupling efficiency of 80% was stably realized under the cavity was filled with gaseous para-hydrogen at a density of 1 x $10^{19}$ cm$^{-3}$.

In conclusion, here we have reported, for the first time to our knowledge, the development and characteristics of a DFIL cw laser. The laser system consists of a Ti:sapphire laser as a power oscillator, two diode lasers employed as seed lasers, and a master Fabry-Perot cavity providing a frequency reference. Stable DFIL oscillation was achieved with a maximum output power of 2.8 W. As fundamental performance characteristics of this laser system, we have shown that both the longitudinal and transverse modes have good single-mode characteristics (spectral purity: nearly same as those of the seed lasers; $M^2$: ~1.0), together with practical power stability (fluctuation: less than 2%). Furthermore, as an advanced characteristics, we have demonstrated that the two frequencies can be arbitrarily selected although the power oscillator has a discrete longitudinal-mode spacing. Moreover, their output-power ratios can be controlled precisely over a wide dynamic range of greater than one order by simply manipulating the seed-power ratio at the two frequencies.

Further extension of the development of the present DFIL laser, such as extension toward a 1-Hz-class ultranarrow-linewidth laser by employing an ultrahigh-finesse reference cavity, or extension to a multi-frequency laser by employing more than three seed lasers, is an attractive proposition and is potentially achievable.

**Funding.** Grant-in-Aid for Scientific Research(A) No. 24244065.
**Acknowledgment.** The authors thank K. Nakagawa and S. Uetake for their useful advices. T. G. acknowledges support of a Japanese Government (Monbukagakusho) scholarship.